\documentstyle[11pt]{article}

\newcommand{\be}{\begin{equation}}
\newcommand{\ee}{\end{equation}}
\newcommand{\bea}{\begin{array}}
\newcommand{\ea}{\end{array}}
\newcommand{\beqa}{\begin{eqnarray}}
\newcommand{\eeqa}{\end{eqnarray}}
\newcommand{\bean}{\begin{eqnarray*}}
\newcommand{\eean}{\end{eqnarray*}}
\newcommand{\eqn}[1]{(\ref{#1})}

\def\up#1{\leavevmode \raise.16ex\hbox{#1}}

\newcommand{\journal}[4]{{\sl #1 }{\bf #2} \up(19#3\up) #4}

\newcommand{\gapproxeq}{\lower .7ex\hbox{$\;\stackrel{\textstyle >}{\sim}\;$}}
\newcommand{\lapproxeq}{\lower .7ex\hbox{$\;\stackrel{\textstyle <}{\sim}\;$}}

\newcounter{appendice}

\def\thebibliography#1{{\bf REFERENCES\markboth
 {REFERENCES}{REFERENCES}}\list
 {[\arabic{enumi}]}{\settowidth\labelwidth{[#1]}\leftmargin\labelwidth
 \advance\leftmargin\labelsep
 \usecounter{enumi}}
 \def\newblock{\hskip .11em plus .33em minus -.07em}
 \sloppy
 \sfcode`\.=1000\relax}

\begin{document}

\title{\hfill $\mbox{\small{
$\stackrel{\rm\textstyle DSF-29/97\quad}
{\rm\textstyle UAHEP 9710\quad}
{\rm\textstyle hep-th/9706190\quad\quad}
$}}$ \\[1truecm]
2+1 Einstein Gravity as a Deformed Chern-Simons Theory}
\author{G. Bimonte$^{a}$, R. Musto$^{a}$, A. Stern$^{b}$ and
P. Vitale $^{a}$}
\maketitle
\thispagestyle{empty}

\begin{center}
{\it a)  Dipartimento di Scienze Fisiche, Universit\`a di Napoli,\\
Mostra d'Oltremare, Pad.19, I-80125, Napoli, Italy; \\
INFN, Sezione di Napoli, Napoli, ITALY.\\
\small e-mail: \tt bimonte,musto,vitale@napoli.infn.it } \\
{\it b) Department of Physics, University of Alabama,\\
Tuscaloosa, AL 35487, USA.\\
\small e-mail: \tt astern@ua1vm.ua.edu }\\
\end{center}

\begin{abstract}

The usual description of $2+1$ dimensional Einstein gravity as a
 Chern-Simons (CS) theory is extended to a one parameter
family  of descriptions    of $2+1$ Einstein gravity.
This is done by replacing the Poincar\'e gauge group symmetry
by a $q-$deformed Poincar\'e gauge group symmetry, with the former
 recovered when $q\rightarrow 1$.   As a result, we obtain
 a  one parameter family of
 Hamiltonian formulations for $2+1$ gravity.
Although formulated in terms of noncommuting dreibeins and
spin-connection fields, our
expression for the action and our field equations, appropriately ordered,
are identical in form to the ordinary ones.
Moreover, starting with a properly defined metric
tensor, the usual metric theory can be built; the Christoffel symbols
and space-time curvature  having the usual expressions in terms of
the metric tensor, and being represented by  $c-$numbers.  In this
article, we also
 couple the theory to particle sources, and find that these sources
carry {\it exotic} angular momentum.  Finally, problems related
to the introduction of a cosmological constant are discussed.

\end{abstract}

\bigskip

\newpage

\section*{Introduction}

In $2+1$ dimensions, as shown by E. Witten \cite{wit} (also see
\cite{at}), general relativity is equivalent to a Poincar\'e group gauge
theory  using   a pure Chern-Simons (CS) action. The CS construction has
among the following features:   It is possible only in three dimensions,
due in part,  to the  nondegenerate scalar product on the  Poincar\'e
algebra, which exists only in three dimensions.   On shell, the gauge
symmetries contain  diffeomorphisms of the space-time manifold $M$. From
the CS action one can easily read off the form of the Poisson brackets
and use them to show that the constraints, which state that the field
strengths vanish on a time slice of  $M$,  generate gauge
transformations. The physical phase space of the theory is the space of
flat connections, modulo gauge transformations. The theory is topological
in nature, the only observables being global gauge invariant quantities,
such as holonomies around non-contractible loops of the space manifold,
and edge states if boundaries are present. \cite{bgs}

In this article we show that the equivalence between general relativity
in $2+1$ dimensions and CS theory persists if we replace the  $2+1$
Poincar\'e group by a  $2+1$  {\it quantum}   Poincar\'e group, which we
denote by  $ISO_q(2,1)$, with the usual description  recovered when
$q\rightarrow 1$. Furthermore, we claim that all of the above mentioned
features of the CS theory also persist upon making this replacement.
{\it We thus end up with a one parameter family of descriptions of
Einstein's
general relativity, which exhibits a hidden quantum group gauge symmetry
and a noncommutative structure.}   Concerning     the
 noncommutative structure, although
we regard the space-time manifold $M$ as a commuting manifold, i.e. it is
parametrized by commuting coordinates $x^\mu$, the fundamental fields of
our theory defined on $M$ are noncommuting.   These  fundamental fields
are the analogues of the  dreibeins and components of the spin
connections of Einstein-Cartan theory. Although these fields do not
commute, we can construct out of them a  symmetric
space-time metric, Christoffel symbols and
Riemann curvature tensor, all of which  commute amongst themselves.
While the space-time
metric does not commute with the dreibeins and the spin connections,
the Christoffel symbols and the Riemann curvature tensor, being in
the center of the algebra, can be thought to be ordinary numbers and
hence a sensible gravity theory can be defined.
In a future article\cite
{prep2},  we shall exhibit a similar hidden quantum group gauge symmetry
in four dimensional gravity, although it will obviously  not be described
by a CS theory.

In writing down the theory in $2+1$ dimensions  we shall  utilize the
{\it deformed} CS action, which was constructed  in ref. \cite{bmsv}.
Three  ingredients were found necessary before defining the   deformed CS
theory.  One is the existence of a bicovariant calculus on the quantum
group \cite{woro}.
This allows for a gauge theory description
based on quantum groups, \`a la Castellani \cite{cast1}.  Another
ingredient is a
technical requirement called `minimality' \cite{cast1},\cite{bmsv}, while
the final ingredient is    the existence of a nondegenerate invariant
scalar product, which allows us to write down the action.

All three   of these ingredients are in fact present for $ISO_q(2,1)$. In
ref. \cite{cast2},  it was shown that a bicovariant calculus exists for
certain quantum Poincar\'e groups in any number of dimensions. These
quantum Poincar\'e groups have the feature that they contain the
(undeformed) Lorentz group.   For the case of three space-time
dimensions, we shall give a heuristic  construction of the bicovariant
calculus on $ISO_q(2,1)$ in Section 1.  This requires that we first
define a consistent differential calculus on  $ISO_q(2,1)$, and then show
that it admits commuting left and right transformations, as well as left
(or right) invariant one forms on the quantum group.  From the
bicovariant calculus on $ISO_q(2,1)$, we can then construct the
corresponding gauge  theory.  This is done  in Section 2, and there we
show that it satisfies the minimality requirement.

In Section 3  we give  the  invariant scalar product, and then  write
down the CS action.    As in the undeformed theory, the nondegenerate
scalar product is unique to three dimensions. The equations of motion
following from the CS action take the usual form, i.e., they state that
the analogues of the Lorentz curvature ${\cal R}^a$ and torsion ${\cal
T}^a$ vanish.  However, unlike in Einstein-Cartan theory,    ${\cal R}^a$
and  ${\cal T}^a$, like the dreibeins and the components of the spin
connections, are noncommuting. Nevertheless, as stated above, from these
quantities we can construct mutually  commuting  space-time metrics ${\tt
g}_{\mu\nu}$,  Christoffel symbols $    \Gamma^\sigma_{\mu\nu}$ and
Riemann curvatures  ${{\tt R}^\mu}_{\nu\rho\sigma}$ .  This will be done
in Section 4.  There we also show that the previous equations of motion
imply that, as usual, $  \Gamma^\sigma_{\mu\nu}$   is symmetric in the
lower indices and  ${{\tt R}^\mu}_{\nu\rho\sigma}$ is zero (provided
inverse dreibeins exist).  We thus recover the Einstein equations in the
absence of matter.

We introduce matter in the form of point particles in Section 5. The point
particles in general carry a momentum and angular momentum, the former
defined as the charge for  ${\cal R}^a$ while the latter is a charge for
${\cal T}^a$. The momentum components commute amongst themselves, so in
the absence of any sources for the torsion, we get the usual Einstein
equation coupled to matter.  On the other hand, for consistency,  the
angular momentum components do not commute amongst themselves, nor do
they commute   with the components of momentum,   and we therefore
conclude  that we have particles with exotic spin.  As is usual, from the
Bianchi identities we can obtain   a set of equations of motion for the
particle degrees of freedom.  In Section 6, we show how these equations, as
well as the field equations,  can  be obtained from an action
principle.  It is written explicitly in terms of   $ISO_q(2,1)$
variables.

The CS description of $2+1$ gravity, based on    $ISO(2,1)$ or
$ISO_q(2,1)$ variables applies only in the case of zero cosmological
constant.  In \cite{wit},  Witten showed that $2+1$  gravity with a
nonzero cosmological constant can also be described in terms of a CS
theory, with the gauge group now being $SO(3,1)\, [SO(2,2)]$ for positive
[negative] values of the cosmological constant. In Section 7,  we show that
the CS construction  can be applied to the case of a local {\it
q}-deformed symmetry group $SO_q(3,1)$ or $SO_q(2,2)$. This yields to the
usual expression for the
 Einstein Lagrangian in terms of dreibeins and spin connections
 when a cosmological constant is present,
and to the usual equations of motion. However, we find
that  the contraction of $SO_q(3,1)$ and $SO_q(2,2)$ to $ISO_q(2,1)$ is
singular, unless we first   take the limit $q\rightarrow 1$.  Thus this
theory  is not continuously connected to  the deformed CS gravity with
zero cosmological constant.
Moreover, a symmetric and Lorentz invariant space-time metric cannot be
constructed out of the dreibeins, making an important difference with the
zero cosmological constant theory.

Concluding remarks are given in Section 8.

\section{Bicovariant Differential Calculus on the Quantum Poincar\'e
Group}
\setcounter{equation}{0}
In this Section we first define the relevant $2+1$ Poincar\'e group,
$ISO_q(2,1)$. We then
 construct a differential calculus on the   quantum
group  and show that it admits commuting left and right
transformations, as well as left (or right) invariant one forms.
The results are in agreement with \cite{cast2} in the sense that they
yield an equivalent braiding matrix, as we will see in the next
Section, and the same quantum Lie algebra (apart from a rescaling of the
generators).

The $2+1$  quantum  Poincar\'e group will be expressed in terms of
matrices $\ell= [\ell_{ab}]$ and vectors  $z= [z_{a}]$.  Roman letters
in the beginning of the alphabet indicate Lorentz indices.  We   shall
raise and lower them using    the {\it off-diagonal} Lorentz metric
tensor  $\eta_{ab}$
\be
\eta=\pmatrix{ & & 1\cr & 1 & \cr 1 & & \cr}\;.\label{3dmet}
\ee
We find it convenient to choose them
 to take values $a,b,c,d...=-1,0,1$.   Then we have
 that $\eta_{ab}= \delta_{a+b, 0}$.

Unlike  for the ordinary Poincar\'e group, the matrix elements
$\ell_{ab}$ and  $z_{a}$ are not c-numbers.   Our choice shall be such
 that the  $\ell_{ab}$'s  commute amongst   themselves and the
$z_{a}$'s commute amongst themselves, but that they  don't  commute
with each other.  Instead, we take
\be
z^a {\ell_c}^{b}  =  q^b \;  {\ell_c}^{b} z^a \;,\label{cz}
\ee
where we  introduce the real  deformation parameter $q$. No further
commutation relations are needed to define the universal enveloping
algebra generated by   $\ell_{ab}$ and  $z_{a}$.

We note further that
$\ell_{ab}{\ell_c}^b $ is in the center of the algebra generated by
${\ell_c}^{d}$ and $z^b$.  From (\ref{cz})
\be
z^e  \;{\ell_{a}}^d{\ell_c}^b \eta_{db} =
q^{d+b}{\ell_{a}}^d{\ell_c}^b \eta_{db}  \;
z^e    ={\ell_{a}}^g{\ell_c}^b \eta_{gb} \;
 z^e  \;,
\ee
since $\eta_{db} =\delta_{d+b, 0}$.   Therefore we may set
\be
\ell_{ab}{\ell_c}^b =\eta_{ac} \;,\label{dolm}
\ee
and
the commutation relations (\ref{cz}) which we have chosen  are
consistent with  ${\ell_c}^{d}$ being a Lorentz matrix.
Unlike with $\ell_{ab}{\ell_c}^b$, $\ell^{ba}{\ell_b}^c$
(where we now sum on the first index in $\ell$)
 is not in the center of the algebra generated by
${\ell_f}^g$ and $z^f$ except for certain values of $b$ and $c$. From
\eqn{cz}
\be
z^e\ell^{ba}{\ell_b}^c = q^{a+c} \ell^{ba} {\ell_b}^c z^e
\ee
Nevertheless, we may still set
 \be
\ell^{ba}{\ell_b}^c= \eta^{ac}~,   \label{dolm2}
\ee
as is done for the Lorentz group, since in so doing we are putting all
noncentral components of $\ell^{ba}{\ell_b}^c$ equal to zero.

Also as is
done for the Lorentz group, the determinant of $\ell$ may be set
equal to one, as it is in the center of the algebra generated by
${\ell_f}^{g}$ and $z^f$.   For this we need to introduce  the totally
antisymmetric tensor ${\cal E}_{abc}$,  with $ {\cal E}_{-1\;0\;1} =
1$.  Now
\be
{\rm det} (\ell)\;   {\cal E}_{def} \equiv
{\cal E}_{abc} {\ell_d}^a {\ell_e}^b  {\ell_f}^c
\ee
commutes with $z^g$ due to the identity
\be
q^{a+b+c}\;  {\cal E}_{abc}=  {\cal E}_{abc}  \;.
\ee
Hence we can set
\be
{\rm det} (\ell) =1 \;. \label{det1}
\ee

Our $2+1$ quantum  Poincar\'e group is  defined as
the universal enveloping
algebra generated by    ${\ell_c}^{d}$ and $z^b$, subject to the
constraints (\ref{dolm}),  (\ref{dolm2}) and (\ref{det1}).
These constraints imply that $ISO_q(2,1)$ contains the (undeformed)
Lorentz group.

We next wish to construct a differential calculus on the space spanned
by   $\ell_{ab}$ and  $z_{a}$.  For this  we must specify the
commutation relations for   $\ell_{ab}$ and  $z_{a}$ with their
exterior derivatives, the choice being consistent with (\ref{cz}).  A
natural choice is
\beqa
d z^a {\ell_c}^{b} & =  &    q^b   \;
 {\ell_c}^{b} dz^a \cr
  z^a d{\ell_c}^{b} & =  &    q^b  \;d
 {\ell_c}^{b} z^a \cr
d z^a \wedge d{\ell_c}^{b} & =  & -   q^b  \;d
{\ell_c}^{b} \wedge d z^a \;,  \label{crawl}
\eeqa
while assuming that the calculus on the space generated by
  $\ell_{ab}$ alone is the usual one on $SO(2,1)$, i.e.
\beqa
d{\ell_a}^b {\ell_c}^{d} & =  & {\ell_c}^{d} d {\ell_a}^b \cr
d{\ell_a}^b \wedge d{\ell_c}^{d} & =  &  -
d {\ell_c}^{d}\wedge d {\ell_a}^b \;,     \label{crall}
\eeqa
and the calculus on the space generated by
  $z_{a}$ alone is the usual one on ${\bf R}^3$, i.e.
\beqa
d   z^b z^{d} & =  & z^{d} dz^b \cr d   z^b \wedge d z^{d} & =
 & -dz^{d}\wedge d z^b \;.\label{zz}
\eeqa

Left and right transformations can be introduced in the usual way,
i.e. as if the variables  $ \ell $ and $z  $  were standard Poincar\'e
group matrices. For the case of infinitesimal {\it left} transformations,
variations of ${\ell_c}^{d}$ and $z^b$ take the form:
\be
\delta_L {\ell}^{ba}= -{{\cal E}_{dc}}^b    \tau_L^d  {\ell}^{ca}
\;,\qquad \delta_L z^b =-  {{\cal E}_{dc}}^b \tau_L^d z^c +
\rho_L^b \;, \label{lft}
\ee
where $ \tau_L^b$ and $ \rho_L^b $  are infinitesimal parameters.
${\cal E}_{ab}^{\;\;\;\;c}$ are  $so(2,1)$ structure constants
\be
{\cal E}_{ab}^{\;\;\;\;c}= {\cal E}_{abd}\; \eta^{cd}\;.
\ee
   Our choice of indices gives ${\cal
E}_{ab}^{\;\;\;\;c}= {\cal E}_{a  b\;-c}  $, and furthermore the
result that    ${\cal E}_{ab}^{\;\;\;\;c}$ is  nonzero only for
$a+b=c$.    In addition, we have the usual identity
\be
{\cal E}_{ab}^{\;\;\;\;c}    {\cal E}_{dec} =-\eta_{ad}\eta_{be}
+\eta_{ae}\eta_{bd } \label{usid} \;.
\ee
The  transformations (\ref{lft}) differ  from the usual left
Poincar\'e transformations because the infinitesimal parameters are
not  c-numbers.  For this we notice that the commutation relations
(\ref{cz}),  (\ref{crawl}), (\ref{crall}) and (\ref{zz}) are preserved
under left transformations provided that
\be
[\rho_L^a,{\ell^{c}}_b]_{q^{b}}=0\;, \quad [\rho_L^a,z_b]=0\;,
\ee
where we use the notation $[A,B]_s=AB-sBA$, and $ \tau_L^b$
commutes with   ${\ell_c}^{d}$ and $z^b$. Similar  commutation
relations are assumed for $ \tau_L^b$ and $ \rho_L^b $  with the
one forms  $d{\ell_c}^{d}$ and $dz^b$.

For the case of infinitesimal {\it right} transformations, variations of
${\ell_c}^{d}$ and $z^b$ take the form:
\be
\delta_R {\ell_c}^{d}=  {\ell_c}^{f}  {{\cal E}_{fe}}^d \tau^e_R
\;,\qquad \delta_R z_b =  \ell_{ba} \rho_{R}^a \;, \label{rht}
\ee
where $ \tau_R^b$ and $ \rho_R^b $  are infinitesimal parameters.
This transformation differs from the usual right  Poincar\'e
transformations by the fact that the infinitesimal parameters are not
c-numbers.  For this we notice that the commutation relations
(\ref{cz}),  (\ref{crawl}), (\ref{crall}) and (\ref{zz}) are preserved
under left transformations provided that
\be
\left. \matrix{[\tau^a_R,\ell^{bc}]=0   \;, &
  [\tau_R^a,z^b]  _{q^{-a}}=0\;,\cr
  [\rho^a_R,\ell^{cb}]_{q^{b}}=0\;,&
[\rho^{a}_R,z^b]_{q^{-a}}=0\;,\cr } \right.\label{crlar}
  \ee
along with similar  commutation relations with the one forms
$d{\ell_c}^{d}$ and $dz^b$. It can be easily checked that the
infinitesimal left and
right transformations (\ref{lft}) and (\ref{rht})
commute (at the second order in the variations).

To construct a bimodule it remains to write down the left (or right)
invariant one forms.  The left invariant forms  are given by the usual
expressions for the Poincar\'e group, i.e
\be
\omega ^c _L= \frac12  {\cal E}_{ab}^{\;\;\;\;c}
(\ell^{-1} d\ell)^{ab}  \;,\quad
 e ^c_L =(\ell^{-1} dz)^{c}  \;,   \label{drso}
\ee
although here the ordering in the expression for $e_L ^c$ is crucial.
>From (\ref{cz}), $(\ell^{-1} dz)^{c}= {\ell_b}^c dz^b$ $= q^{-c}\;
dz^b{\ell_b}^c$.  It is straightforward to check that (\ref{drso}) is
invariant under global left transformations (\ref{lft}).    (By global
we mean that $d\tau_L^b=d\rho_L^b =0$.)
 Alternatively, the right
invariant one forms are given by the usual  expressions for the
Poincar\'e group, i.e
\be
\omega_R^c = \frac12  {\cal E}_{ab}^{\;\;\;\;c}
(d\ell \ell^{-1} )^{ab}  \;,\quad
 e_R^c ={\ell^c}_b d(\ell^{-1} z)^{b}  \;.
\ee

In what follows,  however, we shall work exclusively in terms of the
left invariant forms  (\ref{drso}).  For convenience we shall drop
the $L$ subscripts and just denote      the
left invariant forms  by $\omega^a$ and $e^a$.
They transform nontrivially  under right transformations.
Since from now on we shall  deal with right transformations only, in the
rest of the paper we shall
omit writing the subscript $R$ on the infinitesimal gauge parameters
and just denote them by $\tau^a$ and $\rho^a$.  We then have the
following expressions for  the variations of the
left invariant forms   (\ref{drso})    under
 infinitesimal right transformations   (\ref{rht}):
\beqa
\delta_R \omega^c & =&  d\tau^c  +
{\cal E}_{ab}^{\;\;\;\;c} \omega^a \tau^b\;,\cr
\delta_R e^c & = & d\rho^c  +{\cal E}_{ab}^{\;\;\;\;c}(
\omega^a \rho^b -   \tau^a e^b)\;. \label{tras}
\eeqa
>From   the commutation properties (\ref{crlar}), we get
\be
\left. \matrix{[\tau^a,\omega^b]=0   \;, &   [\tau^a,e^b]
_{q^{-a}}=0\;,\cr   [\rho^a,\omega^b]_{q^{b}}=0\;,&
[\rho^{a},e^b]_{q^{b-a}}=0\;.\cr } \right.\label{lroe}
\ee
The  left invariant forms (\ref{drso}) satisfy the Maurer-Cartan
equations
\be
{\cal R}^c   =   {\cal T}^c   =0    \;,\label{RTez}
\ee
where
\beqa
{\cal R}^c & =& d\omega^c+{1\over 2} {\cal E}_{ab}^{\;\;\;\;c}
\omega^a\wedge \omega^b\;, \cr {\cal T}^c &  =& de^c+   {\cal
E}_{ab}^{\;\;\;\;c}  \omega^a\wedge e^b\;.\label{dRT}
\eeqa
$  {\cal R}^c $ and ${\cal T}^c$ will denote the spin curvature and
torsion, respectively.

\section{$iso_q(2,1)$ Gauge Theory}
\setcounter{equation}{0}
>From the bicovariant calculus of the preceding Section, we can  write
down a gauge theory based on the $iso_q(2,1)$ algebra.  The algebra
is described in terms of $q$-structure constants and braiding matrices.
In this Section
we shall identify these quantities and verify that they satisfy the
necessary identities defining a minimal quantum group gauge theory.
\cite{cast1}\cite{bmsv}

In defining the quantum group gauge theory, the
right transformations (\ref{tras}) are to be regarded as gauge
transformations on connection one forms $\omega^a$ and $e^a$ [where
now we no longer  assume that they are pure gauges, as in eq.
(\ref{drso}), and thus eq. (\ref{RTez}) no longer applies].  Although
the one forms have nonstandard commutation properties, the underlying
space-time manifold $M$ on which they are defined can be  assumed to
be an ordinary manifold,  parametrized by coordinates.

The gauge transformations (\ref{tras}) can be expressed compactly by
\be
\delta A^i= d\epsilon^i  +C^i_{jk} A^j \epsilon^k \;,\label{ivai}
\ee
where the roman letters starting from the middle of the alphabet are
$iso_q(2,1)$ indices.   We choose them  to take values
$i,j,k,\ell...=-1,0,1,...,4$. We identify $\omega^a =  A^a$ and  $e^a
= A^{a+3}$. To make the connection with gravity, the former denote
spin connections, while the latter are  dreibein one forms. Similarly,
we split the infinitesimal parameters $\epsilon^i$ into infinitesimal
Lorentz transformations $\tau^a = \epsilon^a$ and translations
$\rho^a = \epsilon^{a+3}$.  $ C^k_{ij}$ denote the q-structure
constants for  $iso_q(2,1)$. From (\ref{tras}),  we can make the
following identifications:
\be
C^c_{ab} =   C^{c+3}_{a\;b+3} = {\cal E}_{ab}^{\;\;\;\;c} \;,
\quad  C^{c+3}_{a+3\;\;b} =  q^{-b} \;
{\cal E}_{ab}^{\;\;\;\;c} \;,\label{dsc}
\ee
with all other  components  equal to zero.   In the limit
$q\rightarrow 1$,  we recover the structure constants for $iso(2,1)$.

The commutation properties (\ref{lroe}) can be expressed compactly by
\be
\epsilon^i   A^j=  \Lambda^{ij}_{k\ell}  A^k
\epsilon^\ell\;,\label{Aep}
\ee
where  $\Lambda_{ij}^{k\ell}$ are the
components of the braiding matrix. From (\ref{lroe}) we can make the
identifications
 \be
\Lambda_{ab}^{cd}=\delta^d_a \; \delta^c_b \;,\quad
\Lambda_{a\;\;b+3}^{c+3\;\;d}=q^a\;  \delta^d_a  \;\delta^c_b \;,\quad
 \Lambda_{a+3\;\; b}^{c\;\;d+3}=q^{-b}\; \delta^d_a\;\delta^c_b
\;,\quad  \Lambda_{a+3\;\;b+3}^{c+3\;\;d+3}=   q^{a-b}\;\delta^d_a\;
\delta^c_b \;,
\ee
with all other components equal to zero. In the limit $q\rightarrow
1$, $\Lambda_{ij}^{k\ell}\rightarrow \delta_{i}^{\ell}\delta_{j}^{k}$.

For arbitrary $q$, the braiding matrix satisfies the Yang-Baxter
equation
\be
\Lambda_{k\ell}^{ij}\Lambda_{sp}^{\ell m} \Lambda_{qu}^{ks}=
\Lambda_{k\ell}^{jm}\Lambda_{qs}^{ik} \Lambda_{up}^{s\ell} \;, \label{yb}
\ee
as well as
\be\Lambda_{ij}^{k\ell} \Lambda_{k\ell}^{mn} =\delta^m_i \delta^n_j \;.
\label{mindef}
\ee
 We note that the  q-structure constants are not in general
antisymmetric in the lower two indices.  Instead they satisfy
\be
C^k_{ij} = - \Lambda_{ij}^{rs} C^r_{rs} \;,\label{skewsym}
\ee
along with the q-Jacobi equations:
\be
C^r_{mi}  C^n_{rj} -  \Lambda_{ij}^{k\ell} C^r_{mk}C^n_{r\ell} -
C^k_{ij}  C^n_{mk} =0\;,  \label{qjac}
\ee
as well as a couple of other useful identities
\beqa
\Lambda_{mk}^{ir} \Lambda_{n\ell}^{ks} C^j_{rs} &  =&
 \Lambda_{k\ell}^{ij} C^k_{mn} \;,\cr
\Lambda_{rs}^{jq} \Lambda_{k\ell}^{si} C^r_{ps} +
 \Lambda_{pi}^{jq} C^i_{k\ell}& =&
\Lambda_{r\ell}^{sq} \Lambda_{pk}^{ir} C^j_{is} +
 \Lambda_{pk}^{jr} C^q_{r\ell} \label{2d}  \;.
\eeqa

Using the braiding matrix and q-structure constants, we can construct
the $iso_q(2,1)$ algebra.   It is expressed in terms of generators
$T_i$,  $i=-1,0,1,...,4$, according to
\be
T_iT_j-\Lambda_{ij}^{k\ell} T_kT_\ell =C^k_{ij} T_k\;.\label{qcom}
\ee
We define $T_a= J_a$ and $T_{a+3} = P_a$, the former being analogous
to angular momentum generators, while  the latter are analogous to
momentum generators.  In terms of them (\ref{qcom}) becomes
\beqa
 [J_a,J_b]& =& {\cal E}_{ab}^{\;\;\;\;c} J_c    \;. \cr
[  J_a  ,  P_b]_{q^{a}}&  = &{\cal E}_{ab}^{\;\;\;\;c}  P_c\;,\cr
[P_{a} , P_b]_{q^{a-b}} &  = & 0\;.\label{PP}
\eeqa
The $J-P$ commutation relations can also be written as
\be
[P_a,J_b]_{q^{-b}}=q^{-b} {\cal E}_{ab}^{\;\;\;\;c} P_c\;.
\ee
In our basis, ${1\over\sqrt{2}}(J_{-1} - J_1)$ can be regarded as the
rotation generator, while  ${1\over\sqrt{2}}(J_{-1} + J_1)$ and $J_0$
are boost  generators.  In the limit $q\rightarrow 1$, one recovers
the standard  Poincar\'e algebra in three dimensions, where
${1\over\sqrt{2}}(P_{-1} - P_1)$ can be interpreted as the energy, or
time translation generator, while  ${1\over\sqrt{2}}(P_{-1} + P_1)$
and $P_0$ are space-like momenta, or  spatial translation generators.

The relations (\ref{qcom}), (\ref{yb}), (\ref{qjac}) and (\ref{2d})
define a bicovariant    calculus on the quantum group, while
(\ref{mindef}) is the condition for minimal deformations.   The latter
condition was utilized in finding a consistent Chern-Simons theory
\cite{bmsv}. It
was also needed for the closure of gauge  transformations.\cite{cast1}
Finally, the identity (\ref{skewsym}) follows from (\ref{qcom}) and
(\ref{mindef}). To see this just multiply both sides of (\ref{qcom})
by $\Lambda^{ij}_{rs}$.

If we introduce the q-Lie algebra valued connections $A\equiv A^i
T_i$, then infinitesimal gauge transformations \eqn{ivai} take the familiar
form,
\be
\delta A= d\epsilon +A \epsilon - \epsilon A\;,\label{var1}
\ee
where $\epsilon=\epsilon^i T_i$ and we have used \eqn{Aep}.
>From the commutation relations   (\ref{cz}),  (\ref{crawl}),
(\ref{crall}) and (\ref{zz}) for $\ell$ and $z$, we can compute
the commutation relations for the spin connections and dreibeins using
(\ref{drso}). We get
\beqa
\omega^a\wedge \omega^b&=&-\omega^b\wedge\omega^a~,\cr
e^a\wedge \omega^b &=&-q^b\omega^b\wedge e^a~,\cr
e^a\wedge e^b& =& -q^{b-a}e^b\wedge e^a~.  \label{aa2}
\eeqa
This can be expressed compactly by
\be
A^i\wedge A^j=-\Lambda^{ij}_{k\ell}A^k\wedge  A^\ell \;,\label{aa}
\ee
 which is consistent with   minimal gauge
theories \cite{cast1},\cite{bmsv}.  Due to (\ref{mindef}), no further
conditions arise on $A^i\wedge  A^j$   when  multiplying both sides by
 $\Lambda_{ij}^{k\ell}$.

Due to the commutation relations \eqn{aa}, \eqn{qcom} the curvature two form
\be
F=dA +A \wedge A \;,
\ee
is a $q$--Lie algebra element $F=F^iT_i$, the components $F^i$ being
given by
\be
F^k=dA^k +{1\over 2} C^k_{ij} A^i \wedge A^j \;.
\ee
We can decompose  the latter into the Lorentz curvatures  ${\cal R}^a
=F^a$ and torsions  ${\cal T}^a =F^{a+3}$ given in (\ref{dRT}).
Infinitesimal gauge transformations of $F$ take the familiar form,
\be
\delta F= F \epsilon - \epsilon F\;. \label{dF}
\ee
We say that $F$ transforms under the adjoint action of $ISO_q(2,1)$.
In terms of  ${\cal R}^a$ and   ${\cal T}^a$ we get
\beqa
\delta_R {\cal R}^c & =&
{\cal E}_{ab}^{\;\;\;\;c}{\cal R}^a \tau^b\;,\cr
\delta_R {\cal T}^c & = & {\cal E}_{ab}^{\;\;\;\;c}(
{\cal R}^a \rho^b -   \tau^a {\cal T}^b)
\;. \label{trasRT}
\eeqa
The commutation properties of $F^i$ with $\epsilon^j$ are assumed to be
\be
\epsilon^i   F^j=
 \Lambda^{ij}_{k\ell}  F^k \epsilon^\ell\;,\label{Fep}
\ee
which can be expanded to
\be
\left. \matrix{[\tau^a,{\cal R}^b]=0   \;, &
  [\tau^a,{\cal T}^b]  _{q^{-a}}=0\;,\cr
  [\rho^a,{\cal R}^b]_{q^{b}}=0\;,&
[\rho^{a},{\cal T}^b]_{q^{b-a}}=0\;.\cr } \right.\label{lrRT}
\ee
We should also specify the commutation relations of the curvature two
forms $F^i$ among themselves and with the connections. They have to be
consistent with \eqn{aa}. We assume
\be
A^i\wedge F^j=\Lambda^{ij}_{k\ell}F^k\wedge  A^\ell \;,
\ee
and
\be
F^i\wedge F^j=\Lambda^{ij}_{k\ell}F^k\wedge  F^\ell \;,
\ee
which in terms of the Lorentz curvatures and torsions are
\beqa
\omega^a\wedge {\cal R}^b & =& {\cal R}^b\wedge  \omega^a \;,\cr
   \omega^a\wedge {\cal T}^b & =&  q^{-a}  \;
  {\cal T}^b\wedge  \omega^a \;,\cr
e^a\wedge {\cal R}^b & =&  q^b   \;
  {\cal R}^b\wedge  e^a \;,\cr
e^a\wedge {\cal T}^b & =&  q^{b -a} \;
{\cal T}^b\wedge e^a \;,\label{RToe}
\eeqa
and
\beqa
{\cal R}^a\wedge {\cal R}^b &=& {\cal R}^b\wedge  {\cal R}^a \;,\cr
{\cal R}^a\wedge {\cal T}^b & =&  q^{-a}   \;
{\cal T}^b\wedge  {\cal R}^a \;,\cr
{\cal T}^a\wedge {\cal T}^b & =&  q^{b -a} \;
{\cal T}^b\wedge {\cal T}^a \;,\label{RTRT}
\eeqa
respectively.

The Bianchi identities for the theory can be expressed in the usual
way according to
\be
dF + A\wedge F - F\wedge A=0 \;,
\ee
or, in terms of the spin curvature and torsion, by
\beqa
d{\cal R}^c & =& {\cal E}_{ab}^{\;\;\;\;c}
 {\cal R}^a\wedge \omega^b\;, \cr
d{\cal T}^c &  =& {\cal E}_{ab}^{\;\;\;\;c} (
 {\cal R}^a\wedge e^b - \omega^a\wedge {\cal T}^b)\;.\label{bieo}
\eeqa
Once again the ordering is crucial.

\section{Deformed Chern-Simons Gravity}
\setcounter{equation}{0}
We now utilize \cite{bmsv} and write down a CS action for the quantum
group gauge theory of Section 2.   For this we
specialize to the case where the space-time manifold  $M$ is
three dimensional.

  In order to write down  the Chern-Simons theory,
we will need to specify an invariant q-group metric $g_{ij}$.
Invariance for minimal gauge groups means that \cite{bmsv}
\be
g_{\ell k} C^\ell_{ij} =
 g_{i\ell} C^\ell_{jk}\;.\label{scalinv2}
\ee
This condition is satisfied for the following choice of $g_{ij}$:
\be
g_{a\;3-a}= 1\;,\quad g_{3- a\;a}= q^{-a}\;,\quad ({\mbox{no~sum~ on~}}a)
\label{ndm}
\ee
with all other  components equal to zero. Here we note that the
q-group metric is not symmetric.  Upon using the notation  of ref.
\cite{bmsv}, $g_{ij}=<T_i,T_j>$, we have that
$$
<J_a,P_b>=  \eta_{ab}\;,\quad <P_a,J_b>=    q^{a}\eta_{ab}\;,
$$
\be
<J_a,J_b>=  <P_a,P_b>=  0\;.
\ee

The Chern-Simons Lagrangian density found in \cite{bmsv} is given by
\beqa
{\cal L}_{CS}& =& <dA +{2\over 3} A^2 ,A> \cr
& =& g_{ij} \; (dA^i + {1\over 3} C_{k\ell}^i A^k\wedge  A^\ell
   )\wedge  A^j \;. \label{lag}
\eeqa
For minimally deformed gauge theories it was shown to be gauge
invariant (up to a total derivative), and to have its exterior
product equal to $<F,F>$. Upon using (\ref{dsc}) and (\ref{RToe})
we are able to write (\ref{lag}) as follows:
\be
{\cal L}_{CS}=2 (d\omega^c+{1\over 2} {\cal E}_{ab}^{\;\;\;\;c}
\omega^a\wedge \omega^b)\wedge e_c \;+\;q^a d(e^a \wedge \omega_a) \;,
\ee
where $e_a=\eta_{ab} e^b =e^{-a}$, while its exterior derivative is
given by
\be
<F,F> = 2 {\cal R}^c  \wedge  {\cal T}_c \;,
\ee
using (\ref{RTRT}). If the three manifold $M$ has no boundary then
the associated  Chern-Simons action is just
\be
{\cal S}_{CS}=\int_M
{\cal L}_{CS}=\int_M2 {\cal R}^c \wedge e_c \;,\label{qact}
\ee
which is the usual expression for Chern-Simons gravity\cite{wit},
although once again here the ordering is crucial. Under gauge
transformations (\ref{tras}) and (\ref{trasRT}),
\be
\delta(  {\cal R}^c  \wedge e_c)  =
d(  {\cal R}_c \rho^ c)\;,
\ee
and hence the action ${\cal S}_{CS}$ is gauge invariant.

The equation of motion  ${\cal R}^c = 0 $ follows immediately  from
extremizing the action with respect to variations in  the dreibeins.
In varying the spin connection, let us assume, as usual, that  the one
forms $\omega^a$ anticommute with their  variations, i.e.
\be
\delta\omega^a\wedge \omega^b=-
 \omega^b\wedge  \delta\omega^a \;,
\ee
This is consistent with the variations $\delta \omega^a$ being gauge
variations  $\delta_R \omega^a$ as in   (\ref{tras}). Then after
integrating by parts,
\beqa
\delta {\cal S}_{CS}& =& \int_M 2\delta\omega^a
\wedge  ( d e_a +  {\cal E}_{ab}^{\;\;\;\;c}
\omega^b \wedge e_c ) \cr  & =& \int_M 2
\delta\omega^a \wedge  {\cal T}_a  \;,
\eeqa
yielding ${\cal T}_a = 0$.

As expected, the equations of motion associated with  ${\cal L}_{CS}$
state that the  $iso_q(2,1)$  curvature vanishes, i.e. we recover
eq. (\ref{RTez}).  On simply connected space-time manifolds $M$,
we then get vacuum solutions of the form  (\ref{drso}).

>From ref. \cite{bmsv} we know that $F=0$  follows  from
(\ref{lag}), provided that the   matrix $H_{ij} =\Lambda_{ij}^{k\ell}
g_{k\ell}+ g_{ij}$ is nondegenerate, which is the case for our
choice of q-group metric tensor $g_{ij}$.   An alternative metric
tensor  exists which is instead degenerate,  leading also to a
degenerate $H_{ij}$. It is  simply the Cartan-Killing metric for the
$so(2,1)$ Lie-subalgebra:
\be
g^{(CK)}_{ab}=\eta_{ab}\;,\label{CKm}
\ee
with all remaining components  $g^{(CK)}_{ij}$ equal to zero.    It is
easy to show that (\ref{CKm}) satisfies the  invariance condition
(\ref{scalinv2}).  [More generally,  any linear combination of
(\ref{ndm})  and (\ref{CKm}) satisfies the invariance condition.]
>From this metric tensor  we get the standard $so(2,1)$ Chern-Simons
Lagrangian
 \be
{\cal L'}_{CS}= (d\omega^c + {1\over 3}
{\cal E}_{ab}^{\;\;\;\;c}
   \omega^a\wedge  \omega^b)\wedge  \omega_c \;,\label{cswdm}
\ee
whose associated  equations of motion are just ${\cal R}^c = 0 $.
Thus from ${\cal L'}_{CS}$ we obtain no conditions on the torsion.
Furthermore, the classical dynamics is preserved if we add the term
(\ref{cswdm}) to  ${\cal L}_{CS}$,  although the resulting Poisson
structure will not be.

Because the action (\ref{qact}) (and also the one obtained from $
{\cal L'}_{CS}$) does not require a space-time metric (although it
does require the Lorentz metric $\eta_{a b}$ and the q-group metric
$g_{ij}$ and/or $ g^{(CK)}_{ab}$) , it is, as usual, invariant under
diffeomorphisms,   and  thus topological. Infinitesimal
diffeomorphisms of the connection one form are given by
\be
 \delta_\xi A ={\cal L}_\xi A = i_\xi dA   + d i_\xi A\;,\label{idtr}
\ee
where $\xi$ is an infinitesimal vector field, and ${\cal L}_\xi$ and $
i_\xi $ denote, respectively,   the Lie derivative and contraction
with  $\xi$. Eq. (\ref{idtr})  can be reexpressed in terms of the
curvature two form $F$ according to
\be
 \delta_\xi A =  i_\xi F   + d i_\xi A+ A i_\xi A
  -i_\xi A A \;,
\ee
which is in the form of an infinitesimal  gauge transformation after
imposing the equations of motion $F=0$ and setting $\epsilon =i_\xi
A$. In terms of the spin connections and dreibeins, the latter implies
$\tau ^a =\xi^\mu \omega ^a_\mu$ and $\rho^a =\xi^\mu e^a_\mu$. The
fact that the two symmetry generators coincide is the reason why there
aren't two distinct sets of symmetry generators, one for the gauge
symmetry and one for  diffeomorphisms, in the Hamiltonian formulation
of the theory\cite{bmsv}. This result has been noted by many authors
\cite{Jack} for undeformed Chern-Simons theory.  The novelty  here is
that the components  of $\epsilon$ and  of $ A$ are q-numbers,
although we assume, as usual, that those of $\xi$ are ordinary
c-numbers.  Furthermore, from (\ref{Aep}) and (\ref{aa}), the
commutation properties of $\epsilon$ and $i_\xi A$ are identical.

We now turn to the canonical formalism. We thus consider the action (\ref{lag})
on a three manifold $M=\Sigma \times R$, where $\Sigma$ is a two-manifold
playing the role of space, while $R$ accounts for time.
In ref.\cite{bmsv} it was shown how
to write consistent equal time Poisson brackets for the space components
of the q-CS
field for a generic q-deformed CS theory and we refer the reader to that paper
for the details. Here we just write the result; the only
nonvanishing Poisson brackets are:
$$
\{e^a_i (\bar x), \omega^b_j (\bar y)\}=-q^{b}
\{\omega^b_j (\bar y), e^a_i (\bar x)\}= \eta^{ab}\epsilon_{ij}
\delta^2(\bar x - \bar y),~~~
$$
\be
i,j=1,2~~~\epsilon_{12}=-\epsilon_{21}=1~,  \label{PB}
\ee
where $\bar x$ and $\bar y$ are points of $\Sigma$.  Notice that the q-Poisson
 brackets are not antisymmetric: the extra factor of $q^b$ that arises
after the interchange of the arguments
reflects the commutation relations \eqn{aa2} and is
a consequence of the
fact that the invariant metric (\ref{ndm}) on the q-Poincar\'e algebra
is not symmetric. The time components of the q-CS field play the role of
Lagrange multipliers for the constraints
\be
{\cal T}^a_{ij}\approx 0~,~~~{\cal R}^a_{ij}\approx 0~.
\ee
In ref.\cite{bmsv} it is shown that they are closed with respect to
the Poisson brackets (\ref{PB}), and that   they generate
time-independent gauge transformations, namely local Poincar\'e
transformations on $\Sigma$.

\section{Recovering $2+1$ Einstein Gravity}
\setcounter{equation}{0}
Here we show that from the noncommuting generators of the deformed
CS theory, we can define a space-time metric $
{\tt g}_{\mu\nu}$, Christoffel symbols
 $ \Gamma^\sigma_{\mu\nu} $    and a Riemann curvature
 ${{\tt R}^\mu}_{\nu\rho\sigma}$,
  all of which mutually commute.   Furthermore, from
the equations of motion $F=0$ of the previous Section, we get that
 $ \Gamma^\sigma_{\mu\nu} $  is symmetric in the lower two indices and
 ${{\tt R}^\mu}_{\nu\rho\sigma}$      vanishes, provided  `inverse'
dreibeins exist.     In addition, we end
up with the usual expression for
 ${{\tt R}^\mu}_{\nu\rho\sigma}$ and
 $ \Gamma^\sigma_{\mu\nu} $ in terms of ${\tt g}_{\mu\nu}$.

We first introduce
the space-time metric on $M$.  Its definition, however, is not unique.
 If we define  it as is standardly done by
\be
  \eta_{ab}\;
e^a_\mu e^b_\nu    =    e^{-1}_\mu e^1_\nu +
e^{0}_\mu e^0_\nu + e^{1}_\mu e^{-1}_\nu \;,\label{stm}
\ee
($e^a_\mu$ denoting  the space-time components of the dreibein one
form $e^a$ ; $\mu$ and $\nu$ being space-time indices) then it, like
the q-group metric, is not symmetric. To show this, we need the
commutation relations between different space-time components of the
connection one forms $A^i$.
Since the space--time manifold is described by ordinary commuting
coordinates we may replace \eqn{aa2} by the stronger conditions
\beqa
\omega^a_\mu \omega^b_\nu & =&
  \omega^b_\nu  \omega^a_\mu \;,\cr
e^a_\mu \omega^b_\nu & =&  q^b   \;   \omega^b_\nu  e^a_\mu \;,\cr
e^a_\mu e^b_\nu & =&  q^{b -a} \; e^b_\nu e^a_\mu \;.\label{crstc}
\eeqa
Then (\ref{stm}) is equal to
\be
q^2  e^{1}_\nu e^{-1}_\mu +
e^{0}_\nu e^0_\mu + q^{-2} e^{-1}_\nu e^{1}_\mu \;,
\ee
and hence it is not symmetric. Moreover (\ref{stm}) is not invariant under
local Lorentz transformations, as we would like it to be.

Alternatively, a  symmetric  space-time
metric  $ {\tt g}_{\mu\nu}$, which is also invariant under local Lorentz
transformations,  can be introduced on $M$ by deforming the
usual definition (\ref{stm})  to
\be
{\tt g}_{\mu\nu} =q^{{a-b}\over 2}
\;  \eta_{ab} \; e^a_\mu e^b_\nu    \label{symmet} \;.
\ee
The ordering of the dreibein components is crucial. Although   $ {\tt
g}_{\mu\nu}$ is symmetric, the tensor elements are not c-numbers. From
(\ref{crstc}), we get that   different components of ${\tt
g}_{\mu\nu}$ do not commute with the $iso_q(2,1)$ connection one forms
$\omega^a$ and $e^a$,
\beqa
{\tt g}_{\mu\nu} \;\omega^a_\rho & =&  q^{2a}\;
\omega^a_\rho\; {\tt g}_{\mu\nu}   \;,\cr
{\tt g}_{\mu\nu}\; e^a_\rho & =&  q^{2a}
 \;   e^a_\rho\; {\tt g}_{\mu\nu}    \;.
\eeqa
Nevertheless, components of ${\tt g}_{\mu\nu}$ do  commute with
themselves.

By substituting the vacuum  solution (\ref{drso}) into the definition
(\ref{symmet})  of the symmetrized metric tensor we get the usual
result
\beqa
{\tt g}_{\mu\nu} & =& q^{a}\; \eta_{ab} \;  {\ell_d}^z \partial_
\mu z^d \;{\ell_{e}}^b\partial_\nu z^e   \cr
 & =& q^{a + b}\;\eta_{ab}  \;
{\ell_d}^z{\ell _{e}}^b   \;
\partial_\mu z^d  \partial_\nu a^e   \cr
&=&\eta_{de}\partial_\mu z^d  \partial_\nu z^e\;.
\eeqa
(Remember though that the $z$'s in the above equation are not c-numbers,
as they do not commute with the $\ell$'s, see eq.(\ref{cz}).)
A local Lorentz frame is then obtained by choosing the commuting
functions $z^a$ equal to the space-time coordinates of the manifold
$M$.

Next we write  the Christoffel symbols $   \Gamma^\sigma_{\mu\nu} $
in terms of the  (symmetrized) space-time metric tensor.  We introduce
  the Christoffel symbols along with the spin connections in the
definition of the covariant derivative, and demand, as is usual, that
the covariant derivative of the dreibein components vanish, i.e. ${\tt
D}_\mu e^b_\nu = 0$, where
\be
{\tt D}_\mu e^b_\nu =  {\cal D}_\mu e^b_\nu
 -  \Gamma^\sigma_{\mu\nu} e^b_\sigma \;, \label{cddr}
\ee
and
\be
{\cal D}_\mu e^b_\nu =\partial
_\mu e^b_\nu + {\cal E}_{cd}^{\;\;\;\;b}\omega_\mu^c e^ d  _\nu \;.
\ee
The torsion being zero is consistent with   the Christoffel symbols
being  symmetric in the lower two indices. Upon assuming
\beqa
\partial_\rho e^a_\mu \omega^b_\nu & =&  q^b   \;
\omega^b_\nu \partial_\rho e^a_\mu \;,\cr
\partial_\rho  e^a_\mu e^b_\nu & =&  q^{b -a} \;
 e^b_\nu  \; \partial_\rho  e^a_\mu \;,
\eeqa
along with (\ref{crstc}), we deduce from (\ref{cddr}) that   the
Christoffel symbols commute with the dreibeins and spin connections,
and then consequently, also with the space-time metric and with
themselves.

>From ${\tt D}_\mu e^b_\nu = 0$,  we  can eliminate the spin
connections, if we multiply on the left    by $q^{a} \eta_{ab}
e^a_\rho$, sum over the $b$ index, and symmetrize with respect to the
space-time indices $\nu$ and $\rho$.  The result is
\be
0=q^{a} \eta_{ab}[ e^a_\rho
\partial _{\mu} e^b_{\nu} +
 e^a_\nu \partial _{\mu} e^b_{\rho} -
 e^a_\rho e^b_{\sigma}   \Gamma^\sigma_{\mu\nu}
-e^a_\nu e^b_{\sigma}   \Gamma^\sigma_{\mu\rho} ] \;.
\ee
Next we add to this the equation obtained by switching $\mu$ and
$\nu$, and subtract the equation obtained by replacing indices
($\mu,\nu,\rho)$  by  ($\rho,\mu,\nu)$.  We can then isolate  $
\Gamma^\sigma_{\mu\nu} $ according to
\be
2q^{a} \eta_{ab} e^a_\rho e^b_{\sigma}\Gamma^\sigma_{\mu\nu} =
q^{a} \eta_{ab}    [ e^a_\rho (\partial _{\mu} e^b_{\nu}
+\partial _{\nu} e^b_{\mu})
+ e^a_\nu (\partial _{\mu} e^b_{\rho} -
\partial _{\rho} e^b_{\mu} )
+e^a_\mu (\partial_\nu e^b_{\rho} - \partial_\rho e^b_{\nu}) ]
\ee
or
\be
2  {\tt g}_{\rho\sigma}    \Gamma^\sigma_{\mu\nu}   =
     \partial _{\mu}   {\tt g}_{\rho\nu}
+\partial _{\nu}  {\tt g}_{\rho\mu}-\partial_\rho {\tt g}_{\nu\mu} \;.
\label{chris}
\ee
In order to solve the above equation for the Christoffel symbols, we need
to invert the metric ${\tt g}_{\mu \nu}$.
To do this, we start from defining the inverses $e_a^{\mu}$ of the
co-vectors $e^a_{\mu}$.
 All we need to do is to enlarge our algebra by a new element
 called
${\tt e}^{-1}$ fulfilling the following commutation relations:
$$
{\tt e}^{-1} e^a_{\mu}=q^{-3 a} e^a_{\mu} {\tt e}^{-1} ~~,
$$
\be
{\tt e}^{-1} \omega^a_{\mu}=q^{-3 a} \omega^a_{\mu} {\tt e}^{-1} ~,
\label{einv}
\ee
and such that
\be
{\tt e}^{-1}{\tt  e} =1~,\label{eeinv}
\ee
where ${\tt e}$ is the determinant:
\be
{\tt e}=
\epsilon^{\mu \nu \rho } e^{-1}_{\mu} e^0_{\nu} e^1_{\rho}~,
\label{det}
\ee
and $\epsilon^{\mu \nu \rho } $ is antisymmetric in the space-time indices.
It is easy to verify that eq.(\ref{eeinv}) is consistent, because its
left hand side
commutes with everything. Moreover, it is also true that
 ${\tt e}^{-1} {\tt e} = {\tt e} {\tt e}^{-1}$.
Using ${\tt e}^{-1}$ we can now define the inverses of the dreibeins
 to be:
\be
e^{\mu}_a \equiv -\frac{1}{2}
\hat{{\cal E}}_{a b c }
\epsilon^{\mu \nu \rho} e^b_{\nu} e^c_{\rho}
{\tt e}^{-1} ~,
\ee
where the q-antisymmetric tensor $\hat{{\cal E}}_{a b c }$ is defined such that
\be
\hat{{\cal E}}_{a b c}
e^a \wedge e^b \wedge e^c =
e^{-1} \wedge e^0 \wedge e^1~,~~~{\rm no~sum~on}~a\;, b \;, c~.
\ee
The solution to this equation may be expressed by
$\hat{{\cal E}}_{a b c}$  $= q^{a-c+2} {\cal E}_{a b c} \;.$
It can be checked that $e^{\mu}_a$ satisfy the usual properties:
$$
e^a_{\mu} e^{\mu}_b = e^{\mu}_b e^a_{\mu}  = \delta^a_b~;
$$
\be
e^a_{\nu} e^{\mu}_a = e^{\mu}_a e^a_{\nu}  = \delta^{\mu}_{\nu}~.
\ee
Their commutation properties can be worked out to be:
$$
e^{\mu}_a e^{\nu}_b= q^{b-a} e^{\nu}_b e^{\mu}_a~,
$$
$$
e^{\mu}_a e_{\nu}^b = q^{a-b} e_{\nu}^b e^{\mu}_a~,
$$
\be
e^{\mu}_a \omega ^b_{\nu}= q^{-b} \omega^b_{\nu} e^{\mu}_a~.
\ee
Using the vectors $e^{\mu}_a$ we can now define the inverse of the metric
$
{\tt g}
_{\mu \nu}$ as:
\be
{\tt g}^{\mu \nu} \equiv q^{\frac{a-b}{2}}\eta^{ab} e^{\mu}_a e^{\nu}_b~,
\ee
where $\eta^{ab}$ is the ordinary inverse of the matrix $\eta_{ab}$.
The inverse metric ${\tt g}^{\mu \nu}$ is symmetric and
satisfies the standard conditions:
\be
{\tt g}^{\mu \rho} {\tt g}_{\rho \nu}=
{\tt g}_{\nu \rho} {\tt g}^{\rho \mu}=\delta^{\mu}_{\nu}~.
\ee
Moreover the ${\tt g}^{\mu \nu}$ commute among themselves and with the ${\tt g}
_{\rho \sigma}$ (but do not commute with the dreibeins and with the spin
connection), thus in eqs.(\ref{chris}) we can solve for the
Christoffel symbols without
any ordering problems. The final expression of $\Gamma^{\sigma}_{\mu \nu}$
in terms of the metric and of its inverse coincides with the standard one.

It remains to recover the Einstein equations, which in vacuum state
that the space-time curvature ${{\tt R}^\mu}_{\nu\rho\sigma}$  is
zero.   ${{\tt R}^\mu}_{\nu\rho\sigma}$ is defined by
\be
{{\tt R}^\mu}_{\nu\rho\sigma} v^\nu = 2 ({\tt D}_\rho {\tt  D}_\sigma
- {\tt D}_\sigma {\tt D}_\rho)
 v^\mu \;,
\ee
for any (commuting) space-time vector $v^\mu$. By multiplying by
$e^a_\mu$, we can relate it to the spin curvature
\beqa
e^a_\mu   {{\tt R}^\mu}_{\nu\rho\sigma} v^{\nu}
&  = &  2 ({\tt D}_\rho {\tt  D}_\sigma
- {\tt D}_\sigma {\tt D}_\rho)     e^a_\mu v^\mu    \cr
&  = & 2 ({\cal D}_\rho {\cal D}_\sigma -
{\cal D}_\sigma {\cal D}_\rho) e^a_\mu v^\mu  \cr &=&
{\cal E}_{bc}^{\;\;\;\;a} {\cal R}_{\rho\sigma}^b e^c_\nu v^{\nu}\;.\label{ttcs}
 \eeqa
Thus ${{\tt R}^\mu}_{\nu\rho\sigma}$ is zero as a result of $ {\cal
R}^b $ being zero.     Furthermore, if we replace \eqn{RToe} by the
stronger conditions
\beqa
\omega^a_\mu {\cal R}^b_{\rho\sigma} & =&
  {\cal R}^b_{\rho\sigma}  \omega^a_\mu \;,\cr
e^a_\mu {\cal R}^b_{\rho\sigma} & =&       q^b\;
  {\cal R}^b_{\rho\sigma}  e^a_\mu \;,
\eeqa
then (\ref{ttcs}) implies that the space-time curvature commutes with
everything.

\section{Point Sources}
\setcounter{equation}{0}
Here we include point sources in the field equations in a standard
manner\cite{gs}, i.e. by including delta function contributions to
the field equations.   We presume that a particle traces out a world
line $y^\mu(\tau)$ on $M$, $\tau$ being a real parameter.
We then  endow the particle with `momentum' and `angular momentum'
degrees of freedom, $p^a(\tau)$ and $j^a(\tau)$, respectively.
  Now instead of  (\ref{RTez}), we have                 \beqa
 \frac\kappa 2 \epsilon^{\mu\nu\lambda}  {\cal R}^a_{\nu\lambda} (x)
 &  = &  \int d\tau \;\delta^3(x -y(\tau)) p^a(\tau)
  \partial_{\tau}     {y^\mu} \;,\cr
 \frac\kappa 2 \epsilon^{\mu\nu\lambda}  {\cal T}^a_{\nu\lambda} (x)
 &  = &  \int d\tau \; \delta^3(x -y(\tau)) j^a(\tau)
  \partial_\tau y^\mu\;,\label{ptsou} \eeqa where
$ \partial_{\tau}=  \frac {d}{d\tau}$,
$ \epsilon^{\mu\nu\lambda} $ is totally antisymmetric in the space-time
indices, and we introduce the `gravitational' coupling constant $\kappa$.
Below we give some properties of the particle
degrees of freedom  $p^a(\tau)$ and $j^a(\tau)$.

   From the commutation relations
(\ref{RTRT}),  $p^a(\tau)$ and $j^a(\tau)$ are not c-numbers.  For
consistency with (\ref{RTRT}) we require
 \beqa
[p^a, p^b] \; =  \;  [p^a, j^b]_{q^{-a}}  \;
= \;[j^a,j^b ]_ { q^{b -a}} &=&   0\;,\cr
[\omega^a, p^b] = \; [e^a, p^b]_{q^b} \; =
  \;  [\omega^a, j^b]_{q^{-a}}  \;
= \;[e^a,j^b ]_ { q^{b -a}} &=&   0\;.\label{cpojp}
\eeqa
Thus the particle momentum commutes with itself, but not with the
angular momentum.   It then follows that in the absence of source
terms for the torsion the Einstein-Cartan equations are the usual ones.

Under gauge transformations, $p^a$ and $j^a$ transform as ${\cal R}^a$
and ${\cal T}^a$ (\ref{trasRT}),
\beqa\delta_R p^c & =&
{\cal E}_{ab}^{\;\;\;\;c}p^a \tau^b\;,\cr
\delta_R j^c & = & {\cal E}_{ab}^{\;\;\;\;c}(
p^a \rho^b -   \tau^a j^b)
\;, \label{traspj}         \eeqa    where here
$\rho^b$ and $\tau^a$ denote infinitesimal
functions of the particle's
 space-time coordinates $y^\mu(\tau)$.    From
(\ref{lrRT}), we get the commutation relations between these functions
 and $p^a$ and $j^a$,
\be  \left. \matrix{[\tau^a,p^b]=0   \;, &
  [\tau^a,j^b]  _{q^{-a}}=0\;,\cr
  [\rho^a,p^b]_{q^{b}}=0\;,&
[\rho^{a},j^b]_{q^{b-a}}=0\;.\cr } \right.\label{lrpj}  \;.\ee

By substituting (\ref{ptsou}) into
   the    Bianchi identities (\ref{bieo}),
we get the usual equations of motion for $p^a(\tau)$ and $j^a(\tau)$,
 \beqa \partial_\tau p^c & =& {\cal E}_{ab}^{\;\;\;\;c} p^a
  \omega_\mu^b   \partial_\tau y^\mu \;,\cr
 \partial_\tau j^c & =& {\cal E}_{ab}^{\;\;\;\;c} (
 p^a e^b_\mu - \omega^a_\mu j^b) \partial_\tau y^\mu \;.\label{tpeom}
    \eeqa    From them we find the usual conserved quantities
\be {\cal C}_1= p_a p^a  \quad{\rm and} \quad {\cal C}_2= p_a j^a \;, \ee
the first being analogous to the mass squared, and the second being
 analogous to the mass times spin.
However here, although  ${\cal C}_1$ is in the center of the algebra
generated by  $p^a$ , $j^a$, $e^a$ and $\omega^a$,  ${\cal C}_2$ is not, since
\be [{\cal C}_2, p^b]_{q^b} = [{\cal C}_2, j^b]_{q^b} =  0 \;.\ee
We therefore conclude that particle sources have an exotic spin.

\section{Particle Lagrangian}
\setcounter{equation}{0}
Here we write down an action  for the above mentioned point sources. By
extremizing the particle action along with ${\cal S}_{CS}$ we will
recover both the field equations (\ref{ptsou}), as well as the particle
equations (\ref{tpeom}). In addition, the action is consistent with the
commutation properties (\ref{cpojp}).

Following \cite{bmss},\cite{ss},  we write the momentum and angular
momentum variables in terms of group variables, the group now being
$ISO_q(2,1)$. We therefore again utilize the  matrix elements
$\ell_{ab}$ and  $z_{a}$, only here they are functions on the particle
world line,  $\ell_{ab}=   \ell_{ab}(\tau)$,    $z_{a}=z_{a}(\tau)$. We
shall assume the commutational properties (\ref{cz}).

We now express   $p^a$ and $j^a$ according to
\be
p^a = \ell^{ba} \hat{t}_b \;,\quad     j^a = \ell^{ba} \hat{s}_b +
 {\cal E}^{abc} p_b (\ell^{-1}z)_c \;, \label{pjitolz}
\ee
where $ \hat{t}_b$ and  $ \hat{s}_b $ are to be regarded as constants. In
order for (\ref{pjitolz}) to be consistent with the commutational
properties (\ref{cpojp}) and (\ref{cz}), these constants cannot be
c-numbers.  Instead we can assume that all  the commutation relations
with  $ \hat{s}_b$ and  $\hat{t}^b$ are trivial except for
\be
\hat{s}^a {\ell_c}^{b}  =  q^b \;  {\ell_c}^{b} \hat{s}^a \;.
\label{chs}
\ee
The set of  all $p^a$ and $j^a$ satisfying  (\ref{pjitolz}) defines the
adjoint orbit of $ISO_q(2,1)$. Using  (\ref{pjitolz})  and (\ref{rht}),
$p^a$ and $j^a$ gauge transform according to the adjoint action of
$ISO_q(2,1)$, i.e.  as in (\ref{traspj}).\footnote{Gauge transformations
here differ from those in reference \cite{ss}.  For the latter they
correspond to left transformations of $ISO(2,1)$, while here  they
correspond to right transformations of $ISO_q(2,1)$.}  In other words,
gauge variations of
\be
p^a J_a +j^aP_a \label{pJjP}
\ee
are the same as for $F$ in (\ref{dF}). Upon introducing the covariant
derivative $D_\tau$ of $\ell$ and $z$, we can define another set of
quantities which gauge transform in the same way as (\ref{pJjP}):
\be
\frac12   {\cal E}^{dbc} (\ell^{-1} D_\tau\ell)_{bc} J_d +
    (\ell^{-1} D_\tau z)^d P_d  \;,\label{cdolz}
\ee
where
\be
\frac12  {\cal E}^{abc} (\ell^{-1} D_\tau\ell)_{bc}=  \frac12  {\cal
E}^{abc} (\ell^{-1} \partial_\tau\ell)_{bc}-\omega^a_\tau\;, \quad
(\ell^{-1} D_\tau z)^a=      (\ell^{-1} \partial_\tau z)^a-e^a_\tau\;,
\ee
and
$\omega^a_\tau$   and $e^a_\tau$  are the connection one forms
evaluated on the particle world-line, i.e.
$\omega^a_\tau=\omega^a_\tau (\tau) $  =
$\omega^a_\mu(y (\tau) ) \partial_\tau y^\mu$ and
$e^a_\tau=e^a_\tau (\tau) $  =
$e^a_\mu(y (\tau) ) \partial_\tau y^\mu$.

The scalar product $<\;,\;>$
defined in Section 3 is adjoint invariant.  Hence, we can define a gauge
invariant particle Lagrangian $L_p$ as the scalar product of
 (\ref{pJjP}) and (\ref{cdolz}):
\beqa
L_p&=&< p^a J_a +j^aP_a\;,\;  \frac12
 {\cal E}^{dbc} (\ell^{-1} D_\tau\ell)_{bc} J_d +
(\ell^{-1} D_\tau z)^d P_d> \cr   &=& p_a    (\ell^{-1} D_\tau z)^a
+\frac12 q^{-a} j_a   {\cal E}^{abc} (\ell^{-1} D_\tau\ell)_{bc} \cr
&=& L_0 -p_a      e_\tau^a  -   q^{-a} j_a   \omega_\tau^a \;,
\eeqa
where $L_0$ is the free particle Lagrangian.  Using the identities
(\ref{dolm}) , (\ref{dolm2}) and (\ref{det1}), the latter  can be
expressed according to
\be
L_0 = (z^a \hat{t}^b +    \frac12
 {\cal E}^{abc} \hat{s}_c)
 ( \partial_\tau\ell \ell^{-1})_{ab}\;,
\ee
up to total derivative terms.

It is now straightforward to obtain the field equations (\ref{ptsou}) by
extremizing the action
\be
\frac\kappa{2} {\cal S}_{CS} + \int d\tau L_P
\ee
with respect to variations in the components of dreibein and spin
connection one forms.   Here we also need
\be
\delta\omega^a\wedge {\cal T}^b  =  q^{-a}  \;
  {\cal T}^b\wedge \delta \omega^a \;.
\ee

Getting the particle equations of motion
 (\ref{tpeom}) requires a little more work.
Upon extremizing the particle action $\int d\tau L_P$ with respect to
variations in the Lorentz vector $z^a$ we get
\beqa
\delta \int d\tau L_P&=&\int d\tau \delta z^e \hat{t}^d\biggl\{
(\partial_\tau\ell\ell^{-1})_{ed} -
   {\cal E}_{abc} {\ell_d}^b{\ell_e}^c \omega_\tau^a\biggr\}\cr
&=&-\int d\tau \delta z^e {\ell_e}^c(\partial_\tau p_c -
   {\cal E}_{abc} p^a \omega_\tau^b)\;,
\eeqa
where we integrated by parts and   used
\be
\delta
  z^a {\ell_c}^{b}  =  q^b \;  {\ell_c}^{b}\delta z^a \;.
\ee
The first equation in (\ref{tpeom}) immediately follows from setting this
variation equal to zero.  The remaining equation  in (\ref{tpeom}) can be
recovered by considering right variations of $\ell$, as in (\ref{rht}).
Then
 \be
\delta
 ( \partial_\tau\ell \ell^{-1})_{ab} =
-  {\cal E}_{abc} {\ell^c}_d \partial_\tau\tau^d \;.\ee
Using this and (\ref{traspj}), variations of the particle action take the
form
\beqa
\delta \int d\tau L_P&=&\int d\tau \biggl\{q^{-c} j_c\partial_\tau
\tau^c +  {\cal E}_{abc}(q^{-b}\tau^c j^a \omega_\tau^b - p^b
\tau^c e^a_\tau)\biggr\}   \cr
&=&-\int d\tau q^{-c}\biggl\{\partial_\tau  j_c -
  {\cal E}_{abc}( \omega_\tau^b j^a - p^b  e^a_\tau)\biggr\}\tau^c
\;,
\eeqa
where we integrated by parts and used $[j^b,\tau^a]_{q^a}=0 $ and
$[j^b,\omega^a_\tau]_{q^a}=0 $. The second equation in (\ref{tpeom})
immediately follows from setting this variation equal to zero.

\section{Inclusion of a Cosmological Constant}
\setcounter{equation}{0}
In \cite{wit} it is shown that a cosmological constant can be included
in $2+1$ dimensional gravity in such a way that the theory is
still described by the Chern-Simons action.  Now, however, the gauge
group is not $ISO(2,1)$, but either $SO(2,2)$ or $SO(3,1)$, depending on
the sign of the cosmological constant, $\lambda$.  The algebra of
$ISO(2,1)$ is generalized to
\be
[J_a, J_b]= \epsilon_{abc} J^c~,~~~~[J_a, P_b]= \epsilon_{abc} P^c~,~~~~
[P_a, P_b]= -\lambda\epsilon_{abc} J^c~. \label{so4}
\ee
If $\lambda$ is positive, this can be seen to be the $so(3,1)$ algebra,
while if   $\lambda$ is negative, \eqn{so4} is the $so(2,2)$ algebra.
To see this we can express the algebra in terms of
generators  $J_a$ and $ P'_a={1\over \sqrt{|\lambda|}}P_a$.

The purpose of this Section is to construct a Chern-Simons theory based
on a quantum deSitter or anti-deSitter group. Such q-groups do exist, as
real forms of $SO_q(4)$ \cite{cast96}, and they have an associated
bicovariant calculus and nondegenerate scalar product. However, we find
that their contraction to $ISO_q(2,1)$ ($\lambda\rightarrow 0$) is
singular, unless we take the limit $q\rightarrow 1$ first. Thus the q-CS
gravity with cosmological constant is not continuously connected to  the
q-CS gravity with zero cosmological constant, at least for the
deformation we use.

The minimal, multiparametric deformation of $SO(2n)$ described in
\cite{cast96} becomes, for $2n=4$, a 1-parameter deformation and is given
in terms of six generators
\be
\{J_{-1}^{+}, J_{0}^{+}, J_{1}^{+},  J_{-1}^{-}, J_{0}^{-},
J_{+}^{-} \}\equiv\{T_i\},~~~i=-1,...,4~, \label{ogen}
\ee
obeying the commutation relations
\be
[J_{a}^{\pm},J_{b}^{\pm}]={\cal E}_{abc} J^{{\pm}c},~~~~~
[J_{a}^{+},J_{b}^{-}]_{q^{2ab}} = 0 \label{so22}
\ee
The $\Lambda$ matrix appearing in \eqn{qcom}
can be read directly from \eqn{so22}; we have
\be
\Lambda_{a~b+3}^{c+3~d}=\delta_a^d \delta_b^c q^{2ab}, ~~~~
\Lambda_{a+3~b}^{c~d+3}=\delta_a^d \delta_b^c q^{-2ab}, ~~~~
\Lambda_{ab}^{cd}=
\Lambda_{a+3\;b+3}^{c+3\;d+3}= \delta_a^d \delta_b^c~~
\ee
with all the other components equal to zero. As a real algebra,
\eqn{so22} can be seen to be the $so(2,2)$ algebra, realized  as the
direct sum of two copies of $so(2,1)$, when $q\rightarrow 1$ ($so(3,1)$
can be obtained by taking a complex combination  of  the generators). Also
note that the q-structure constants are the same as the undeformed ones.

The algebra defined by   \eqn{so22} satisfies all the conditions
 for a minimal bicovariant calculus, i.e. eqs.
(\ref{yb})- (\ref{qcom}).  The bicovariant calculus can be defined on the
quantum group $Fun_q(SO(2,1)\otimes SO(2,1))$
generated by the $3\times 3$ matrices $\ell_+$ and $\ell_-$,
both of which satisfy the defining relations for $SO(2,1)$.  Their
matrix elements $\ell_{\pm ab} $ satisfy trivial commutation relations,
except for
\be
{\ell_-}^{ab} {\ell_+}^{cd} = q^{-2bd}
{\ell_+}^{cd}    {\ell_-}^{ab}    \;.
\ee
The differential calculus
on $ Fun_q(SO(2,1)\otimes SO(2,1))$, along with a consistent set
of left and right actions, can be constructed in a straightforward
manner.  From them we get the left invariant one forms
\be  A ^c _\pm= \frac12  {\cal E}_{ab}^{\;\;\;\;c}
(\ell_{\pm}^{-1} d\ell_\pm)^{ab}  \;.\ee
It is clear that the quantum group $ISO_q(2,1)$ defined in Section 1
cannot be obtained
from    $Fun_q(SO(2,1)\otimes SO(2,1))$ by something analogous to
a group contraction.  It also seems reasonable that a similar statement
 can be made for the corresponding q-Lie algebras, and thus that
the corresponding deformed CS theories cannot be continuously
 connected.

We next perform a rotation   from  $(J_+,J_-)$ basis
to the $(J, P')$ basis, which is  needed if one has hopes
of describing gravity;
$ P'$ is the generator $P$ in \eqn{so4} (in the $q\rightarrow 1$ limit)
 rescaled by a factor of $1/\sqrt{|\lambda|}$.   To be definite we assume
from now on $\lambda$ to be negative (hence the gauge algebra to be
$so_q(2,2)$); we have then
\be
J_{a}=J_{a}^{+} + J_{a}^{-}~,~~~~~
P'_{a}=J_{a}^{+} - J_{a}^{-}~. \label{ngn}
\ee
For $q \ne 1$ this change of basis is not uniquely defined,  as we can
consider linear combinations of $J^-, J^+$  with coefficients depending
 on $q$.
Among them, \eqn{ngn} is the simplest one.
We denote by   $V_i, i =-1,4$,
\be V_i=A_i^j T_j \;,  \ee   the
generators $J_a =V_a$ and  $P'_a =V_{a+3}$,
where the matrix $A$ can be read off from \eqn{ngn}.
The new generators obey new deformed commutation relations
\be
V_i V_j-\tilde\Lambda_{ij}^{kl} V_k V_l=\tilde C_{ij}^r V_r
\label{newa}
\ee
with a new braiding matrix, $\tilde \Lambda$ and structure constants
$\tilde C_{ij}^k$, given respectively  by
\be
\tilde \Lambda_{ij}^{kl} = A_i^r A_j^s\Lambda_{rs}^{mn} (A^{-1})_m^k
(A^{-1})_n^l
\ee
and
\be
\tilde  C_{ij}^{k}=A_i^r A_j^s C_{rs}^{m} (A^{-1})_m^k ~.
\ee
The new structure constants are easily seen to be independent from the
deformation parameter. In fact, they are identical to the undeformed ones,
appearing in \eqn{so4}. In particular, they are skewsymmetric in the
lower indices and it can be checked that the skewsymmetry is compatible
with the $\tilde\Lambda$--skewsymmetry \eqn{skewsym}, which is a
characteristic of minimal
deformations. On the other hand the braiding matrix becomes more
complicated than in the old basis. The simple structure $\Lambda_{ij}^{kl}
\propto \delta_i^k\delta_j^l$ is not
preserved; we have instead
\beqa
\tilde\Lambda_{ij}^{kl}&=&\delta_i^l\delta_j^k F(q) \pm(\delta_i^{l+3}
\delta_j^k - \delta_i^l\delta_j^{k+3})G(q) - \delta_i^{l+3}\delta_j^{k+3}
H(q)~,~~ {\mbox{for}}~~ i,j\ne 0,3 \label{ltil1}\\
\tilde\Lambda_{ij}^{kl}&=&\delta_i^l\delta_j^k~,~~ {\mbox{for}} ~~i,j=0,3
\label{ltil2}
\eeqa
where $i+3$ is to be intended as $i+3~ {\mbox {mod}}~ 6$; the plus sign
in \eqn{ltil1} holds for $i=j,j+3$. The functions of $q$ appearing in
\eqn{ltil1} are respectively
\be
F(q)= {1\over 4q^2}(q^2+1)^2~~,~~~~
G(q)= {1\over 4q^2}(q^4-1)~~,~~~~
H(q)= {1\over 4q^2}(q^2-1)^2~~. \label{fq}
\ee
Note that both $G$ and $H$ go to zero for $q\rightarrow 1$, while
$F\rightarrow 1$, so that the deformed algebra \eqn{newa} becomes the
Lie algebra \eqn{so4} up the factor $\lambda$. Restoring the factors
of $\lambda$ in the algebra \eqn{newa} it can be checked that the limit
$\lambda\rightarrow 0$ (no cosmological constant) is singular unless the
limit $q\rightarrow 1$ is previously performed. Thus the deformed theory
with cosmological constant is not continuously connected to the one
with zero cosmological constant described in the previous Sections.

To write down Chern Simons gravity we  need an invariant q--metric
on the deformed algebra, according to \eqn{scalinv2}. Since the
structure constants are undeformed we discover that the metric itself
is undeformed. There are two non--degenerate, invariant metrics on
$so(4)$:
\be
<J_a, J_b>=\eta_{ab}~~,~~~~ <P'_a, P'_b>=\eta_{ab}~~,~~~~
<J_a, P'_b>=0~~,~~~~                     \label{metric1}
\ee
\be
<J_a, J_b>=0~~,~~~~ <P'_a, P'_b>=0~~,~~~~
<J_a, P'_b>= <P'_a,J_b> =\eta_{ab}~~.~~~~           \label{metric2}
\ee
The metric which makes the (undeformed)
CS Lagrangian equal to the Einstein-Cartan
Lagrangian of gravity (with cosmological constant) is the latter, as
pointed out in \cite{wit}. The former gives
indeed a term which, added to the Lagrangian, doesn't change the
equations of motion. Due to the metric being undeformed, the expression
for the  Lagrangian  is the same as the undeformed
one (the equations of motion were already known to be
unchanged as shown in \cite{bmsv}). The deformation is hidden in the
noncommutativity of the connection components.
The CS Lagrangian \eqn{lag} for the present q--group becomes:
\be
{\cal L}_{CS}= [d\omega^c + {1\over 3} {\cal E}_{ab}^{\;\;\;\;c}
(\omega^a \wedge \omega^b + e'^a\wedge e'^b)] \wedge e'_c \;+\; [de'^c +
{1\over 3} {\cal E}_{ab}^{\;\;\;\;c}
(\omega^a\wedge e'^b +e'^a \wedge \omega^b)]
\wedge \omega_c \;,
\ee
where the primed connections are rescaled by a factor of
$\sqrt{ |\lambda|}$. Restoring the factors of $\lambda$ and rescaling the
lagrangian by and overall $\sqrt{|\lambda|}$ we get the undeformed
Einstein--Cartan Lagrangian of $2+1$ gravity
\be
{\cal L}_{CS}= (d\omega^c + {1\over 3} {\cal E}_{ab}^{\;\;\;\;c}
\omega^a \wedge \omega^b ) \wedge e_c \;+\; [de^c +
{1\over 3} {\cal E}_{ab}^{\;\;\;\;c}
(\omega^a\wedge e^b +e^a \wedge \omega^b)]
\wedge \omega_c \;-\; {\lambda\over 3} {\cal E}_{ab}^{\;\;\;\;c}
e^a\wedge e^b \wedge e_c                          \label{cosmlag}
\ee
If we try to reorder the terms containing both the spin connection and
the dreibein in a given order, we have to use \eqn{aa} and \eqn{ltil1},
\eqn{ltil2}, which will introduce a q--dependence  in the Lagrangian.
The equations of motion are, as expected, undeformed, for a particular
ordering of the  connections:
\beqa
{\cal R}^c&=& -{\lambda\over 2} {\cal E}_{ab}^{\;\;\;\;c} e^a \wedge e^b
\cr
{\cal T}^c&=& 0~,
\eeqa
where ${\cal R}$ and ${\cal T}$ are now defined as
\beqa
{\cal R}^c & =& d\omega^c+{1\over 2} {\cal E}_{ab}^{\;\;\;\;c}
\omega^a\wedge \omega^b\;, \cr
{\cal T}^c &  =& de^c +  {1\over 2} {\cal E}_{ab}^{\;\;\;\;c}
(\omega^a\wedge e^b + e^a \wedge \omega^b) \;.\label{dRT2}
\eeqa
Such a theory is invariant under gauge transformations \eqn{ivai}, which
can be explicitly written as
\beqa
\delta e^c&=& d\rho^c + {\cal E}_{ab}^{\;\;\;\;c} (\omega^a\rho^b + e^a\tau^b)\\
\delta \omega^c&=& d\tau^c + {\cal E}_{ab}^{\;\;\;\;c} (\omega^a\tau^b -
\lambda e^a\rho^b)~.
\eeqa
Thus, a deformed theory of Einstein--Cartan gravity with cosmological
constant exists with many of the features of the undeformed one.

Unlike what happens in Section 4 for zero cosmological
constant, here we are not able to give a metric formulation of the
theory. That is to say, it is not possible to construct a metric tensor
out of the dreibeins, which is symmetric, and invariant under local
Lorentz transformations. Also, it is not meaningful for the theory
described above to have pure Lorentz transformations, as the
non--commutativity of the gauge parameters with the connection components
\eqn{Aep} and the structure of the braiding matrix generate terms
containing the Lorentz parameter out of terms which do not contain it.
To illustrate these things let us perform a gauge transformation
of a bilinear in the dreibeins. We have
\be
\delta (e^a \wedge e^b)= [d \rho^a + {\cal E}_{cd}^{\;\;\;\;a}
(\omega^c \rho^d + e^c \tau^d)]\wedge
e^b + e^a \wedge [d \rho^b + {\cal E}_{cd}^{\;\;\;\;b} (\omega^c \rho^d +
e^c \tau^d)];
\ee
to compare the two terms we have to put all the gauge parameters on one side,
for example on the right. Consider the first term of the sum and commute
$\rho^d$ with $e^b$; we get
\be
\delta (e^a \wedge e^b)= [d \rho^a  \wedge e^b -{\cal E}_{cd}^{\;\;\;\;a}
(-{H(q) \over \lambda} \omega^c\wedge \omega^b \tau^d  + ... ]
 + e^a \wedge [d \rho^b + {\cal E}_{cd}^{\;\;\;\;b} (\omega^c \rho^d +
e^c \tau^d)];
\ee
we have obtained a factor containing the spin connection and the gauge
parameter $\tau$, which cannot be cancelled by another factor. Also, the same
term shows that it is not possible to perform
pure Lorentz transformations.
Hence we conclude that a metric tensor cannot be constructed for this
theory, though the  Lagrangian which we have exhibited
is a legitimate deformation of the Einstein--Cartan Lagrangian of
gravity with cosmological constant.

\section{Concluding Remarks.}

The increasing mathematical interest in quantum Lie groups, quantum Lie
algebras and deformed affine Lie   algebras \cite{drinf}, as well as
the r\^ole played by them
in such physical problems as $1+1$ solvable
models, has motivated the construction of $q-$deformed gauge field
theories.  To this end, the construction of
 bicovariant calculi on quantum groups has been extensively studied
leading to the formulation of consistent Yang-Mills theories for different
quantum groups, including the $q-$deformed unitary groups. \cite{are}
A further development, utilized in this article, has been the
construction of deformed  CS theory.  This was applicable
 for minimally deformed groups whose associated quantum Lie
algebras are endowed with a nondegenerate invariant scalar product.

Despite its elegance and formal consistency,
the physical relevance of a
{\it classical} field theory formulated in terms of abstract noncommuting
variables needs to be clarified. The results obtained in this paper may
provide a first step in that direction.    Instead of  constructing new
{\it exotic} theories with such variables, one can search for {\it hidden}
 structures  in well known
field theories. (An analogous result was found within the framework of
the classical    mechanics of a rigid rotor \cite{mar}.)
In the present paper we have done this by
showing that (torsion-free) Einstein
general relativity may be reformulated not only as a Poincar\'e group gauge
theory with pure CS action, but more generally, as a $q-$CS theory  based
on a quantum Poincar\'e gauge group.

>From our theory, one has the possibility of gaining
new insights for  gravity,
 apart from the  possibility of coupling it to
matter  endowed with exotic properties such as those
 described in Section 6.
First, one may notice that the theory is  equipped with a
different canonical formalism for each value of $q$, with the standard
one recovered when $q\rightarrow 1$.
Now if one
were able to quantize the theory, perhaps by extending functional
integration to the case of noncommuting variables taking value in a
quantum Lie algebra, it might be possible to generalize known
 mathematical and physical results of CS theory.  For example, by
  generalizing the results of E. Witten \cite{wit2}, one could
associate  new invariants to a class of knots  by taking the expectation
value of path-ordered exponentials of line integrals
 along closed  space-time curves.   The line integrals  would be
functionals of quantum Lie algebra valued gauge fields.
More generally, a quantum  version of
a $q-$deformed field theory involving two parameters (the deformation
dimensionless quantity $q$ and  Planck's constant) would be
instructive.   One could imagine a scenario where $q$ ends up playing
 the  r$\hat o$le of a regularization parameter.
Finally, the most immediate and physically relevant follow up
 of our work is the search for {\it hidden} deformed structures for
 $3+1$ dimensions Einstein general relativity.
As we have already mentioned, this search has been successful,
and we will report on it in a forthcoming article \cite{prep2}.


\begin{thebibliography}{99}


\bibitem{wit} E. Witten, \journal{Nucl. Phys.}{B311}{88}{46}.

\bibitem{at} A. Ach\'ucarro and P.K. Townsend, \journal{
Phys. Lett.}{B180}{86 }{89}.

\bibitem{bgs}  P. Salomonson, B.-S. Skagerstam and A. Stern,
 \journal{Nucl. Phys.}{B347}{90}{769};
C. Vaz, ``Exact Solutions in Topological Gravity", UCEH-102-92; in
Conf. on Topics in Quantum Gravity, Cincinnati 1992, eds F. Mansouri and
J.J. Scanio: Quantum Gravity and Beyond;
S. Carlip, \journal{Canadian Gen. Rel.}{}{93}{215};
 G. Bimonte, K. S. Gupta and A. Stern, \journal{Int. J. Mod.
Phys.}{A8}{93}{653}.
  M. Hayashi,  \journal {Nucl. Phys.}{B405}{93}{228};
  M. Hayashi,  F. P. Zen, \journal {Prof. Theor. Phys.} {91}{94}{361}.
W. G. Unruh and P. Newbury, \journal{Int. J.  Mod. Phys.}{D3}{94}{131};
K. Ezawa, \journal{Phys. Rev.}{D49}{94}{5211};
A. P. Balachandran, L. Chandar and A. Momen, ``Edge States  in Canonical
Gravity", SU-4240-610, talk given at 17th annual MRST meeting,
 gr-qc 9506006; \journal{Int. J. Mod. Phys.}{A12}{97}{625}.

\bibitem{prep2} G. Bimonte, R. Musto, A. Stern and P. Vitale,
               in preparation.

\bibitem{bmsv} G. Bimonte, R. Musto, A. Stern and P. Vitale, ``Deformed
Chern-Simons Theories", Alabama preprint UAHEP 975 and
Napoli preprint DSF 16/97, hep-th/9704118,
Phys. Lett {\bf B}, in press.

\bibitem{woro} S. L. Woronowicz, \journal{Comm. Math. Phys.}{111}{87}{613};
\journal{ibid.}{122}{89}{125}. 
D. Bernard, \journal{Prog. Theor. Phys.
Suppl}{102}{90}{49};~~D. Bernard and A. Le Clair, \journal{Comm. Math.
Phys.}{142}{91}{99}.
B. Zumino, {\it Proc. Math. Phys X Leipzig}~  Springer--Verlag
(1992);~~ P. Schupp, P. Watts and B. Zumino, \journal{Comm. Math.
Phys.}{157}{93}{305}.
S. Watamura, \journal{Comm. Math. Phys.}{158}{93}{67}.~~
P. Aschieri and L. Castellani,
\journal{Int. J. Mod. Phys.}{A8}{93}{1667}.

\bibitem{cast1} L. Castellani, \journal{Mod. Phys. Lett.}{A9}{94}{2835};
                \journal{Phys. Lett.}{B292}{92}{93};~~
                \journal{Phys. Lett.}{B327}{94}{22}.

\bibitem{cast2} L. Castellani,
\journal{Commun. Math. Phys.}{171}{95}{383}.


\bibitem{Jack} See for example, R. Jackiw, {\it Phys. Rev. Lett.}
{\bf 41} (1978) 1635.

\bibitem{gs} See for example, K. Gupta and A. Stern,
\journal{Phys Rev.}{D44}{91}{2432}.

\bibitem{bmss} A. P. Balachandran, G. Marmo, B. S. Skagerstan and A. Stern,
\journal{Phys. Lett.}{B89}{80}{199}; ``Gauge Symmetries and Fibre
Bundles, Applications to particle Dynamics", Lecture Notes in Physics
{\bf 188}, Springer--verlag, Berlin, 1982.
\bibitem{ss} B. S. Skagerstam and A. Stern, \journal{Int. J. Mod.
Phys.}{A5}{90}{1575}.
\bibitem{cast96} P. Aschieri and L. Castellani,
                   \journal{Int. J. Mod. Phys.}{A11}{96}{4513}.

\bibitem{wit2} E. Witten, \journal{Commun. Math. Phys.}{121}{89}{351}.


\bibitem{drinf} M. Jimbo, \journal{Lett. Math. Phys.}{10}{85}{63}.
V. G. Drinfeld, {\it Quantum Groups} ICM Proc. New York,
Berkeley 1986.
L. D. Faddeev, N. Yu. Reshetikin and L. A. Takhtajan,\journal{Algebra and
Analysis}{1}{87}{178}.
S. Majid, \journal{Int. J. Mod. Phys.}{A5}{90}{1}.

\bibitem{are} J. M. Maillet and F. Nijoff \journal{Phys. Lett.}{B229}{89}{71}.~~
I. Ya. Arefeva and I. V. Volovich,
      \journal{Mod. Phys. Lett}{A6}{91}{893}.~~
      M. Hirayama \journal{Prog. Theor. Phys.}{88}{92}{111}.
A. P. Isaev and Z. Popowicz, \journal{Phys. Lett.}{B281}{92}{271};
  \journal{ibid.}{B307}{93}{353}.~~
T. Brezinski and S. Majid, \journal{Phys. Lett}{B
298}{93}{339};
\journal{Comm. Math. Phys.}{157}{95}{591};
\journal{ibid.}{167}{95}{235};
P. M. Hajac,  ``Strong Connections and $U_q(2)$ YM Theory
on Quantum Principal Bundles" hep-th/9406129.
M. Durdevic,  ``Quantum Principal Bundles and Corresponding
Gauge Theories q-alg/9507021;
\bibitem{sud} A. Sudbery, \journal{Phys. Lett.}{B375}{96}{75}.
P. Watts, ``Toward a $q$--Deformed Standard Model" Preprint
hep-th/9603143.

\bibitem{mar} G. Marmo, A. Simoni and A. Stern,
\journal{Int. J. Mod. Phys.}{A10}{95}{99}.

\end{thebibliography}
\end{document}